UDC  621.311.22  621.311.25

V.A. Kostygin, G.S. Stolyarenko, G.M. Kochetov, A.M. Tugay, V.N. Vashchenko

# Water softening by single-bowl ion exchange filter efficiency estimate and improvement

*National University of Construction and Architecture, Kiev, Ukraine*

*The article presents results of experimental investigations of the water softener in a laboratory installation of uninterruptible countercurrent ion exchange filter, which has a movable layer of ion exchange material. The installation provides for two simultaneous processes: counter ion sorption and regeneration of the sorbent with the processing capability of the sorbent in the regeneration zone by ultrasonic radiation.*

Modern science and technology progress headline is the introduction of designs and technologies which possess high technical and economic parameters, and which primarily minimize energy consumption and material expenses, have a high degree of automation and high reliability.

The results of investigations of the water softening using single-bowl ion exchange reactors of uninterruptible action were presented in earlier papers [1, 2]. In these sources, the results of investigations on two laboratory models - ion exchange reactor of uninterruptible action KOSTOL-1, KOSTOL-2 and on a semi-industrial pilot installation KOSTOL-3 are presented and technical and economic analysis of uninterruptible ion exchange filtration compared with devices running on a fixed layer operating water softening process are given.

Optimization of the process of removing hardness salts of feed water running to the needs of energy sector, using ion exchange resins are urgent practical problems. Standard technological challenge is to reduce the hardness salt content from 10-15 mg.eq./dm$^3$ to 0.1-0.01 mg.eq./dm$^3$ [3].

Ion exchange standard design unit is most often a cylindrical vessel having a feed system of the treated water and the regenerating solution which contains a fixed layer of adsorbent (ion exchange resin) through which the fluid being treated flows [4-7]. The design of such units is simple and it reliable in operation. However, these units have a high flow resistance. If the treated liquid contains suspended particles, then the "cementing" of the filter layer happens and consequently the lack of capacity utilization of the filter take place. Resin regeneration in these filters is performed by either once-through manner, or counter-current manner.

Devices with solid sorbent layer are known [7-14]. Among the countercurrent apparatus with solid sorbent layer there are the simplest devices with a gravitational movement of a sorbent column. In such apparatus the sorbent is fed from the top and the solution below the spent sorbent is derived below by means of an ejector pump or an airlift. Their effectiveness decreases rapidly with increasing diameter because due to preferential channels formed by liquid movement ("parasitic" channels).

To improve performance of the resin layer the extra force is provided to prevent its fluidization. This force is generated by means of the pressure exchanger column and circulation of the spent solution ("hydram").

Semi-uninterruptible countercurrent type devices can be represented by installation of Asahi, which is constantly being modernized [15-20]. The device has a secondary drainage solution distributor and the distributor plate. Main drain tank and dispenser, which are connected with the reaction zone through a ball check valve, are mounted on top. The column is operated intermittently. During the sorption solution is fed through the valve and moves upward through the resin layer and is hydraulically pinned to an upper drainage. Check valve prevents displacement of the resin in the tank. To move the resin solution the feed is stopped and the solution is drained from the column bottom via an additional drainage, after this the valve opens and the entire volume of the column is filled with sorbent. Then the solution supply resumes. The saturated sorbent below the liquid distributor is displaced to the regeneration pressure of the processed fluid column. The disadvantage of the Asahi unit is a large number of controlled valves.

A large number of semi-uninterruptible apparatus was created in the USSR. Sorption pressurized column (SPC) has the simplest design [21-24]. SPC is a tank with solution bottom valve, the upper drainage and storage bin resin. The unit operates as follows: the unit is filled with resin, the solution is filtered through the resin from the bottom up. Periodically feed solution is stopped and resin portion is evacuated by an airlift, the same amount of resin flows from the upper part of the hopper apparatus. Then sorption continues. SPC disadvantage is imperfect distribution system required for complete saturation of the resin. Height of unit is significant (up to 10m), it leads to greater restriction.

For the processing of low-concentration solutions at high performance column was designed [25-27]. The design is very similar to Asahi unit. It is characterized by absence of the solution dispenser in the bottom of the reaction zone and the lower drainage to drain the solution. A treated solution supply and withdrawal of spent resin are alternately mate through one nozzle. Pumping resin is carried out by the airlift in case solution supply stops. This unit is simpler than Asahi reactor from the viewpoint of design and management. But due to lack of distributor solution required height of the working layer of the sorbent should be much greater.

Uninterruptible apparatus or so-called "Pachuc" [28-30]. "Pachuc" is a vertical apparatus equipped with two airlifts. Airlifts are exchanger and stirring the solution and the pumping pulps exchanger solution to a drainage device, in which an ion exchanger returns to the contact zone, and the solution was withdrawn from the apparatus. The volume of "Pachuc" is typically a few tens or hundreds of cubic meters, the residence time of the solution in the machine is 20-60 min. Therefore, to achieve the desired process parameters usually a cascade of such devices is used.

Particularly interesting are installations where all process steps are made in one body. The first attempts to create such systems took place in 1950x [5-7]. All of them were practically unsuccessful, as a simple circuit without valves, which would be shared zone, led to the mixing of the ion exchanger. There are few current designs of monohull uninterruptible ion exchange filters, in which there is a combination of two processes: the actual ions sorption by sorbent and regeneration of the sorbent.

The most known filter of uninterruptible action is Higgins contactor [31]. This unit has many upgrades. All its sections are separated from each other by a large diameter valves, and the resin moves periodically during their opening. In contemporary design, there are two options of Higgins contactors with water filtration in the working sections from top to bottom (classic version) and bottom-up. The contactor is a U-shaped loop, the valves divided into sections: sorption, regeneration and washing of the resin, metering and transporting, washing impurities from the resin and chopped. Since these devices are designed for processing low-concentration water solutions, the sorption section has a substantially larger cross section than the others. In devices for processing of concentrated solutions all the sections have equal diameter. Operation of the unit is rather complicated procedure of opening and closing of valves and periodic movement of the ion exchange material.

The main disadvantage of the sorption of ions is the formation in such apparatus within a stationary resin layer of a so-called parasitic channels through which the fluid moves with the lowest hydraulic losses. It leads to premature breakthrough of ions, which determines the need for an earlier regeneration of ion exchange resins, which in turn causes additional overrun of the apparatus and a regenerating agent. A second disadvantage of this technology is that suspended solids are present in the initial process. The fluid entering the resin produces pollution of the filter layer, which results in increased hydraulic resistance and impaired sorption processes - desorption.

Comparative analysis of the existing technologies of reception of softened water suggests the following conclusions:

- the implementation of a periodical regeneration process does not allow the whole exchange capacity of the ion exchange resin;
- in a fixed bed of sorbent undesirable parasitic channels are produced through which leakage of adsorbed ions occurs;
- mechanical impurities presented in the treated water contaminate the filter layer, which leads to an increase in hydraulic resistance;
- poorly sorbed ions (e.g. sodium) are first to pass through the filter layer (due to diffusion broadening of peaks in a fixed bed of sorbent) in the final form all complicates the flow chart, this is especially true in the facilities obtaining highly demineralized water;
- fixed layer of ion exchanger requires expensive reagent for resin restoration.

The purpose of the research is to study process of water softening on the basis of laboratory model of an uninterruptible action single hull ion exchange reactor KOST-KISI-150 *using ultrasonic radiation in sorbent regeneration zone*, definition of rational design and technological parameters of the single hull filter of uninterruptible action.

We have carried out appropriate laboratory and pilot studies *with a rational number of experiments*.

Treated water quality depends on several parameters in a general relationship:

$$C_f = f(x_1 \ldots x_i \ldots x_n) \qquad (1)$$

where $C_f$ – treated water quality, and $x_i$ – parameters of the described filtering process.

Due to small numbers of independent parameters in the filtration process under investigation, the simple classical enumeration procedure of research model parameter was chosen. The application of a well-known Box-Wilson methodology being more complex in terms of mathematics was not used at this stage.

Tasks for research were set to find empirical functional dependence (1) on time, for the cases of different filters and ultrasound application.

This paper presents the results obtained during the 6 months experiment in 2013 in the Kiev National University of Architecture and Construction on the basis of laboratory installation KOST-KISI-150.

**Experimental Procedure**. Given a goal of the study, two series of experiments on the depletion and recovery of the resin were organized. One series of experiments on the operation of the filter depletion and recovery without ultrasound and a second series of experiments using ultrasound. To carry out research laboratory installation KOST- KISI-150 was used. Schematic diagram of the ion-exchange filter device is shown in Figure 1. Exterior view of KOST- KISI-150 installation is presented in Figure 2.

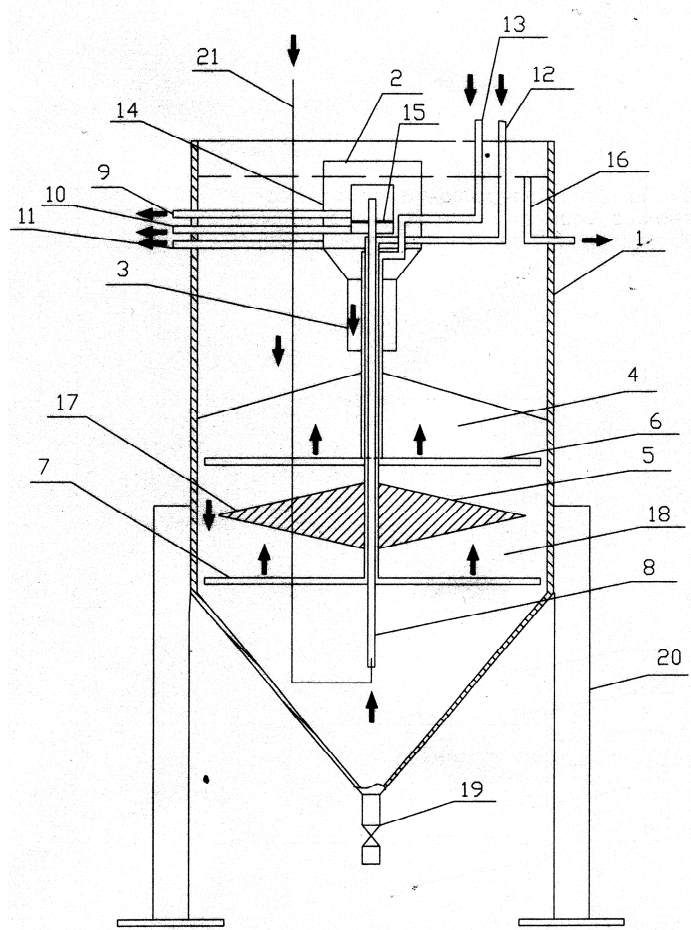

Fig.1. Basic design of KOST-KISI-150 ion exchange filter.
1 - bowl; 2 - clearing unit; 3 - labyrinthiform channel; 4 - operating sorbent layer; 5 – airlock; 6,7 - distribution collectors; 9,10,11- drain connections; 12 - reagent tube; 13 - supply tube; 14 - overflow pipe; 15 – graticule; 16 - drain tube; 17 - annulus; 18 - regeneration zone; 19 – sorbent outlet valve; 20 - filter mount; 21 - airlift.

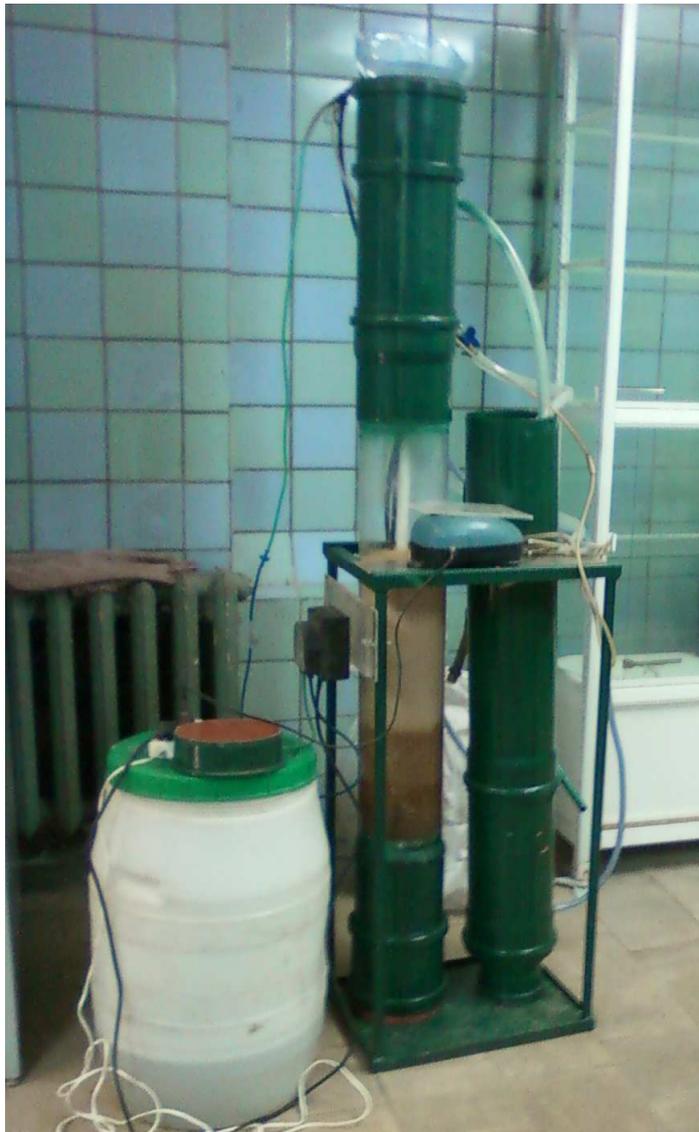

Figure 2. Exterior view of KOST-KISI-150 installation

*Ion exchange filter operation is as follows:*

The treated liquid flows through reagent tube 12 runs to the distribution collector 6, and then to upper operating sorbent layer 4, where hardness ions sorption is made by ion-exchange material. Operating sorbent layer 4 runs countercurrent to the fluid movement, and regenerated sorbent falls on top of operating layer of sorbent from the labyrinthiform channel 3. Exhaust sorbent is supplied to sorbent regeneration zone 18. Treated water comes through the drain tube 16. Regenerating solution is fed via the supply tube assembly 13 through which distribution collector 7 is supplied to the regeneration zone.

The regenerated sorbent by means of airlift 21 runs to overflow pipe 2, a liquid containing desorbed ions and mechanical impurities is discharged through the drainage pipes 9, 10, 11 into drains. Graticule of overflow pipe serves for separating solids and a solution containing desorbed hardness ions. Dehydrated sorbent mass flows over the edge of overflow pipe 15 into labyrinthiform channel 3. As the liquid level in the treated unit is lower than in the bowl of the filter sorbent is washed with purified liquid. The resulting drain fluid is reset through the drain pipe 11. Level difference is created by means connectors 9, 10, 11. An airlock device 17 provides effective mixing of a sorbent and a regenerating solution.

Thus, the countercurrent washing process of regenerated ablution sorbent into labyrinthiform channel 3 is implemented and then the sorbent is supplied to the upper part of the operating sorbent layer. Thus, the process of simultaneous countercurrent sorption hardness ions and regeneration of the sorbent and removing of mechanical impurities which may be in the supplied liquid are implemented.

The main geometrical dimensions of the KOST-KISI-150 filter are: total height - 1.7 m, inner diameter of bowl - 0,150 m, height of the sorbent layer - 1 m. The total maximum flow rate of fed water was $3.9 \cdot 10^{-5} m^3/s$ (140$dm^3$/hour). Water temperature was 15-20°C.

Standard complexometric method was used in the research to determine the overall hardness. It is based on the Regent Trilon B, in accordance with the procedure specified in The State Standard 2874-82 "The Drinking Water". The relative error did not exceed 10-15%. Lower limit of determined concentrations of hardness salts was 0.01 mg.eq/$dm^3$. Total exchange capacity of KU-2-8 cationite produced by Cherkassy PA "AZOT" was 1.8 mmol/g. Maximum capacity for the resin airlift was not higher than 200$cm^3$/min. Source of ultrasonic radiation was a "Reksona" ultrasonic washing machine. Total power consumption was 100W. The machine was placed on the outer side of the lower part of the filter bowl which was in contact with clipper wrapped with a wet sponge.

**Results and discussion**. The obtained results are available in two series. One series of experiments is the work of filter on depletion and recovery without ultrasound and the second series of experiments was made using ultrasound. The most typical results are shown in Tables 1, 2, 3. And Figure 3, 4, 5 show the corresponding graphs.

A first set of experiments was carried out without use of ultrasound at two different speeds of filtration. For A100 series, the filtration rate is 6.3 m/hr, for the A150 series the filtration rate is 8.6 m/hr and for experiments using ultrasound, which was performed with a constant filtration rate, the series are marked U150 with filtration speed 9.9 m/hr.

Table 1

**The results of water treatment at the installation KOST-KISI-150 versus time (A100 series)**

| S No | Intensity of flow dm$^3$/h | Hardness of in-coming water H$_{vch}$ mg – eq/dm$^3$ | Hardness of treated water H$_O$ mg - eq/dm$^3$ | Hardness of concentrated drainage H$_{kd}$, mg - eq/dm$^3$ | Solution NaCl consumption cm$^3$/s | Concentration of solution NaCl % (weight) | Treating time hour |
|---|---|---|---|---|---|---|---|
| 1 | 98,7 | 4,3 | 0,12 | 0.20 | - | - | 1 |
| 2 | 98,7 | 4,3 | 0,16 | 0,20 | - | - | 3 |
| 3 | 98,7 | 4,3 | 0,27 | 0,36 | - | - | 6 |
| 4 | 100,0 | 4,4 | 0,20 | 0,40 | - | - | 7 |
| 5 | 100,0 | 4,4 | 0,30 | 0,50 | - | - | 9 |
| 6 | 100,0 | 4,4 | 0.40 | 0,53 | - | - | 12 |
| 7 | 99,6 | 4,4 | 0,60 | 0,70 | - | - | 13 |
| 8 | 99,6 | 4,4 | 0,70 | 0,70 | - | - | 15 |
| 9 | 99,6 | 4,4 | 0,78 | 1,32 | - | - | 18 |
| 10 | 99,6 | 4,4 | 1,43 | 2,70 | - | - | 20 |
| 11 | 101,3 | 4,4 | 2,42 | 3,40 | - | - | 21 |
| 12 | 101,3 | 4,4 | 4,30 | 6,50 | - | - | 23 |
| 13 | 101,3 | 4,4 | 5,60 | 8,30 | - | - | 26 |
| 14 | 101,3 | 4,4 | 7,20 | 11,30 | 0.32 | 36 | 28 |
| 15 | 100,5 | 4,5 | 6,80 | 11,30 | 0,32 | 36 | 29 |
| 16 | 100,5 | 4,5 | 4,60 | 12,40 | 0,32 | 36 | 31 |
| 17 | 100,5 | 4,5 | 1,70 | 11,50 | 0,32 | 36 | 34 |
| 18 | 100,5 | 4,5 | 0,80 | 11,50 | 0,32 | 36 | 36 |
| 19 | 103,4 | 4,4 | 0,88 | 11.60 | 0,34 | 36 | 37 |
| 20 | 103,4 | 4,4 | 0,55 | 11,70 | 0,34 | 36 | 39 |
| 21 | 103,4 | 4,4 | 0,34 | 11,75 | 0.34 | 36 | 40 |
| 22 | 103,4 | 4,4 | 0,24 | 12,00 | 0,34 | 36 | 41 |

Table 2

**The results of water treatment at the installation KOST- KISI -150 versus time (A150 series)**

| S No | Intensity of flow dm$^3$/h | Hardness of in-coming water H$_{vch}$ mg – eq/dm$^3$ | Hardness of treated water H$_O$ mg - eq/dm$^3$ | Hardness of concentrated drainage H$_{kd}$, mg - eq/dm$^3$ | Solution NaCl consumption cm$^3$/s | Concentration of solution NaCl % (weight) | Treating time hour |
|---|---|---|---|---|---|---|---|
| 1 | 144,2 | 4,5 | 0,10 | 0,20 | - | - | 1 |
| 2 | 144,2 | 4,5 | 0,16 | 0.2 | - | - | 3 |
| 3 | 144,2 | 4,5 | 0,18 | 0,24 | - | - | 5 |
| 4 | 128,7 | 4,4 | 0,10 | 0,50 | - | - | 7 |
| 5 | 128,7 | 4,4 | 0,08 | 0,20 | - | - | 9 |
| 6 | 128,7 | 4,4 | 0,20 | 0,22 | - | - | 11 |
| 7 | 130,0 | 4,6 | 0,60 | 0,58 | - | - | 12 |
| 8 | 130,0 | 4,6 | 0,63 | 0.70 | - | - | 14 |
| 9 | 130,0 | 4,6 | 1,97 | 1,70 | - | - | 16,5 |

| 10 | 130,0 | 4,6 | 2,54 | 2,70 | - | - | 18 |
| 11 | 140,3 | 4,6 | 4,90 | 5.20 | - | - | 19,5 |
| 12 | 140,3 | 4,6 | 8,60 | 13,70 | 0,32 | 36 | 21,5 |
| 13 | 140,3 | 4,6 | 8,40 | 19,00 | 0,32 | 36 | 23,5 |
| 14 | 140,3 | 4,6 | 6,30 | 14,60 | 0,32 | 36 | 25 |
| 15 | 142,7 | 4,5 | 6,10 | 14,40 | 0.32 | 36 | 26 |
| 16 | 142,7 | 4,5 | 4,70 | 18,30 | 0,32 | 36 | 28 |
| 17 | 142,7 | 4,5 | 1,00 | 11,00 | 0,32 | 36 | 29,5 |
| 18 | 142,7 | 4,5 | 0,70 | 11,50 | 0,32 | 36 | 31 |

Table 3

**The results of water treatment at the installation KOST-KISI-150 versus time with ultrasound turned on (U150 series)**

| № п/п | Intensity of flow $dm^3/h$ | Hardness of in-coming water $H_{vch}$ mg – eq/$dm^3$ | Hardness of treated water $H_O$ mg - eq/$dm^3$ | Hardness of concentrated drainage $H_{kd}$, mg - eq/$dm^3$ | Solution NaCl consumption $cm^3/s$ | Concentration of solution NaCl % (weight) | Treating time hour |
|---|---|---|---|---|---|---|---|
| 1 | 157,6 | 4,5 | 3,3 | 4,6 | - | - | 0,5 |
| 2 | 157,6 | 4,5 | 0,55 | 0,40 | - | - | 1,5 |
| 3 | 157,6 | 4,5 | 0,30 | 0,50 | - | - | 3,5 |
| 4 | 157,6 | 4,5 | 0,28 | 0,34 | - | - | 6,5 |
| 5 | 157,3 | 4,6 | 0,26 | 0,31 | - | - | 7,5 |
| 6 | 157,3 | 4,6 | 0,28 | 0,34 | - | - | 9,5 |
| 7 | 157,3 | 4,6 | 0,30 | 0,35 | - | - | 11,5 |
| 8 | 158,0 | 4,6 | 0,30 | 0.34 | - | - | 12,5 |
| 9 | 158,0 | 4,6 | 0,34 | 0,40 | - | - | 14,5 |
| 10 | 158,0 | 4,6 | 0,36 | 0,55 | - | - | 20,5 |
| 11 | 158,0 | 4,6 | 0,60 | 0,70 | - | - | 23,5 |
| 12 | 159,3 | 4,6 | 0,66 | 0,80 | - | - | 24,5 |
| 13 | 159,3 | 4,6 | 0,90 | 0,97 | - | - | 26,5 |
| 14 | 159.3 | 4,6 | 1,20 | 1,40 | - | - | 29,5 |
| 15 | 159,3 | 4,6 | 2,30 | 2,80 | - | - | 31,5 |
| 16 | 159,0 | 4,5 | 3,2 | 4,0 | - | - | 33,5 |
| 17 | 159,0 | 4,5 | 4,0 | 4,3 | - | - | 35,5 |
| 18 | 159,0 | 4,5 | 4,4 | 4,7 | - | - | 37,5 |
| 19 | 159,0 | 4,5 | 4,4 | 6,0 | 0,32 | 36 | 39,5 |
| 20 | 159,2 | 4,5 | 4,3 | 6,2 | 0,32 | 36 | 40,5 |
| 21 | 159,2 | 4,5 | 4,2 | 6,6 | 0,32 | 36 | 42,5 |
| 22 | 159,2 | 4,5 | 3,9 | 7,3 | 0,32 | 36 | 45,5 |
| 23 | 159,2 | 4,5 | 3,6 | 7,3 | 0,32 | 36 | 47,5 |
| 24 | 159,0 | 4,6 | 3,4 | 7,3 | 0,31 | 36 | 49,5 |
| 25 | 159,0 | 4,6 | 2,4 | 7,3 | 0,31 | 36 | 51,5 |
| 26 | 159,0 | 4,6 | 0,90 | 7,3 | 0,31 | 36 | 54,5 |
| 27 | 159,0 | 4,6 | 0,54 | 7,1 | 0,31 | 36 | 56,5 |

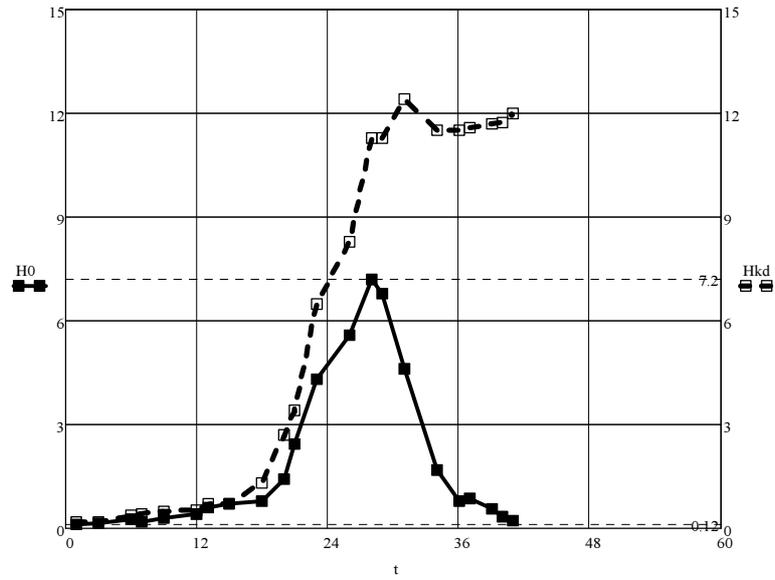

Fig.3. The dependence of the concentration of hardness salts in purified water H0 (mg-eq/dm$^3$) and the concentration of hardness salts in the concentrated drain Hkd (mg-eq/dm$^3$) versus time t (hours). The filtration rate is 6.3 m/hour.

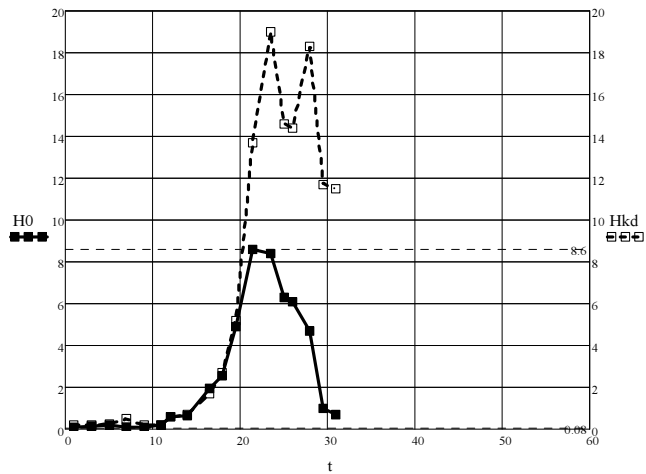

Fig. 4. The dependence of the concentration of hardness salts in purified water H0 (mg-eq/dm$^3$) and the concentration of hardness salts in the concentrated drain Hkd (mg-eq/dm$^3$) versus time t (hours). The filtration rate is 8.6 m/hour.

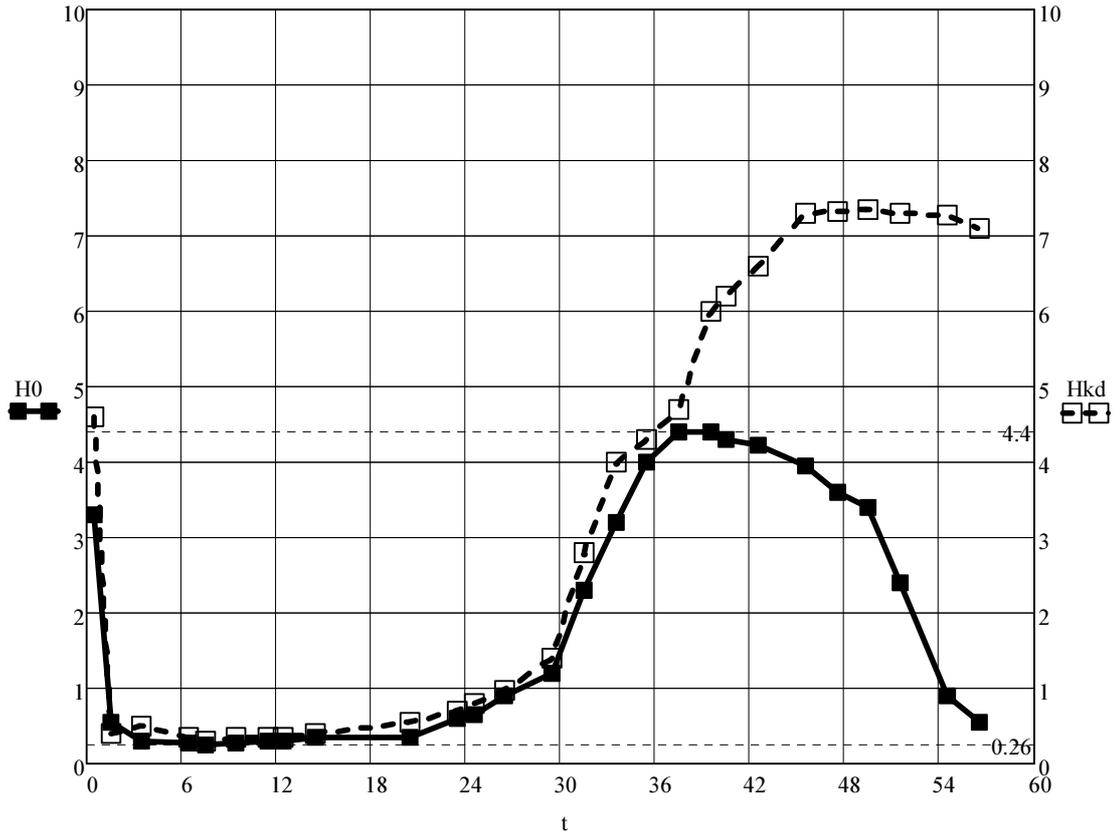

Fig. 5. The dependence of the concentration of hardness salts in purified water H0 (mg-eq/dm$^3$) and the concentration of hardness salts in the concentrated drain Hkd (mg-eq/dm$^3$) versus time t (hour). The filtration rate is 9.9 m/hour. Regenerating zone is treated with ultrasound.

The experimental data obtained indicates that the character of the curve quality obtained with purified water to an ion exchange resin exhaustion interval is well studied characteristic of S-shaped curve, which describes the saturation processes such as enzymatic catalysis, or so-called Michaelis-Menten curve. And in case of conjoint catalysis it is a Hill curve [32]. And finally, in the systems with increasing number of population members it is a so-called logistic curve [33, 34]. General analytical form of this curve is the following:

$$H_{kd} = H_{kdmax}\, e^{k_d(t-z)} /( 1 + e^{k_d(t-z)}) \qquad (2)$$

where $H_{kd}$ – concentration of hardness salts in the drain fluid (mg.eq./dm$^3$);

$H_{kdmax}$ – maximum concentration of hardness salts in the drain fluid (mg.eq./dm$^3$);

Z – delay time (h);

$k_d$ – proportionality constant.

Speaking of the quality of purified water curve with reagent supply we can see that it looks like a falling exponential, usually describing system transformation process to a stationary state.

The solution to this problem in our case is presented in one of the authors' paper [35] and the analytical form of which in our case has the form:

$$H_0 = (H_{00} - \{(H_{00} - H_{0k})\, e^{-(\gamma^* S_f L_{sorb})/Q_f} + H_{0k}\})\, e^{-\beta t} + \\ + (H_{00} - H_{0k})\, e^{-(\gamma^* S_f L_{sorb})/Q_f} + H_{0k} \qquad (3)$$

where $\gamma^*$ - a proportionality factor which depends on the characteristics of the resin and the external physical factors (electromagnetic fields, ultrasound etc. (1/s));

$S_f$ – sectional area of the filter (dm$^2$);
$Q_f$ – productivity of the filter (dm$^3$/s);
$L_{sorb}$ – height of the filtering layer of the filter (dm);
$H_0$ – hardness on the filter outlet (mg-eq/dm$^3$);
$H_{00}$ – hardness on the filter input (mg-eq/dm$^3$);
$H_{0k}$ – stationary rigidity on filter outlet (mg-eq/dm$^3$);
$\beta$ - coefficient depending on the filtration rate, height of the filter layer, the exchange capacity of the ion exchange resin, performance of the airlift (1/s);
 t - time (s).

Given a sufficiently large Lsorb value the formula (3) is converted to the following simple equation:

$$H_0 = (H_{00} - H_{0k})\, e^{-\beta t} + H_{0k} \qquad (4)$$

Analysis of the obtained data suggests the following conclusions:
- the proposed construction of the ion exchange filter enables simultaneously carry out the sorption of hardness ions and regeneration of cation and thus soften the water to the residual hardness of less than 0.1 mg.eq./dm$^3$;
- use of ultrasound allows to use all the exchange capacity of the ion exchange resin, as a result of ion sorption not only in external but also in internal diffusion region of resin beads;
- in a moving layer of sorbent the formation of undesired parasitic channels that prevents breakthrough of adsorbed ions was eliminated;
- mechanical impurities present in the purified water are removed during a constant washing of resin in the labyrinthiform channel, which eliminates the accumulation of the resin layer, and leads to reduction and stabilization of the hydraulic resistances;
- constancy of the concentration gradient of the reagent eliminates the probability of sudden changes in the concentrations, and therefore minimizes the osmotic impact on the cation beads and ultimately reduces the overall number of flawed granules;
- the use of our proposed device also enables extraction of valuable components from waste water, salt solutions, mining and marine waters.